\documentclass[preprint,aps]{revtex4}

\usepackage{epsfig}
\begin{document}
\title{Generic theory of colloidal transport}

\author{Frank J\"ulicher$^1$ and Jacques Prost$^{2,3}$}
\affiliation{$^1$Max-Planck Institute for the Physics of Complex Systems,
N\"othnitzerstr. 38, 01187 Dresden, Germany \\
$^2$ESPCI, 10 rue Vauquelin, 75231 Paris Cedex 05, France\\
$^3$Physicochimie Curie, Institut Curie, 26 rue d'Ulm, 75231 Paris Cedex 05, France}

\begin{abstract}
We discuss the motion of colloidal particles relative to a two component fluid
consisting of solvent and solute. 
Particle motion 
can result from  (i) net body forces on the particle 
due to external fields such as gravity; (ii) slip velocities on the particle surface
due to surface dissipative phenomena. The perturbations of the hydrodynamic
flow field exhibits characteristic differences in cases (i) and (ii) which reflect
different patterns of momentum flux corresponding to the existence
of net forces, force dipoles or force quadrupoles. 
In the absence of external fields, gradients of concentration or pressure do
not generate net forces on a colloidal particle. Such gradients can nevertheless
induce relative motion between particle and
fluid. 
We present a generic description of surface dissipative phenomena based
on the linear response of surface fluxes driven by conjugate surface forces. 
In this framework we discuss different transport scenarios including
self-propulsion via surface slip that is induced by active processes
on the particle surface.
We clarify the nature of force balances in such situations.
\end{abstract}

\maketitle

\section{Introduction}

Motion of colloidal particles immersed in a fluid can be driven by 
gradients of concentration or pressure even in the absence of net body
forces on the particle. 
In such situations, particle motion results from relative motion between
particle and fluid induced by surface slip \cite{ande89}. 
As a result, the particle moves relative to the fluid
without net force.
 
Physical mechanisms that underlie colloidal transport in a fluid
can 
also be used to drive the swimming of self-propelling particles.
Recently, several scenarios have been discussed by which
colloidal objects can self-propel. In the case of self-electrophoresis,
swimming in a fluid is driven by self-generated electric dipole fields,
acting at the fluid particle interface \cite{leon95,lamm96}. 
Phoretic swimmers are driven by
a chemical reaction that is catalyzed in an asymmetric manner on the particle surface 
\cite{gole05,gole07,hows07}.
The surface reaction generates a
concentration field near the particle which is asymmetric. The propulsion
results from surface slip generated by a local concentration gradient at the particle surface. 
Active processes on a surface also
propel many cells and microorganisms.
An important example is motion driven by motile cilia and other active 
cellular processes \cite{bray92}.
 In a coarse-grained description, the beating motion of many cilia on a surface 
effectively generates surface slip velocities that drive propulsion \cite{ligh52,blak71a}.

Here, we discuss a general theoretical framework to describe the motion of
colloidal particles in a systematic and controlled way. Our hydrodynamic description is based on 
conservation laws and linear irreversible thermodynamics in a two-component fluid.
Extension to multicomponent fluids is straightforward.
The basic concepts discussed here are well established \cite{degroot,land87,mart72}.
However, some confusion in the literature calls for clarification \cite{cord08}.

In section II, we review the general hydrodynamic equations of a two-component
fluid. We express the conservation laws for mass, energy and momentum and
identify the conjugate fluxes and forces. The hydrodynamic equations follow from
a description of dissipative fluxes driven by conjugate thermodynamic forces. 
In section III, we discuss
conditions for which concentration gradients and pressure gradients can be 
generated and the time scales during which they persist. A generic theory of dissipative processes
associated with slip velocities on a solid surface that is in contact with a two-component fluid
is discussed in section IV. Section V describes the motion of a colloidal particle
in pressure gradients and concentration gradients. 
Self-propulsion of a colloidal object due to self-generated concentration gradients
as well as due to active surface processes are discussed in section VI.
We conclude our work with a discussion.




\section{Hydrodynamics of a two component fluid}

In order to define all basic concepts clearly, we
first review the hydrodynamics of a two-component fluid
characterized by a hydrodynamic flow field and diffusive fluxes.
Using the systematic formulation of hydrodynamic equations based on 
irreversible thermodynamics \cite{degroot},
we follow the discussion presented in \cite{joan07}.
The fluid consists 
of solvent $a$ and solute $b$ with concentrations (numbers of molecules per unit volume) 
$n_a$ and $n _b$. The molecular
masses of the two components are $m_a$ and $m_b$, respectively.
We consider the incompressible case where the molecular
volumes $v_a$ and $v_b$ are constant. 

{\bf Conservation laws and entropy production}.
The molecular concentrations satisfy
the conservation laws
\begin{equation}
\frac{\partial n_i}{\partial t}+\nabla \cdot {\bf J}_i=0
\end{equation}
with $i=a,b$, where ${\bf J}_i$ denote particle currents.
The mass density of the fluid is 
$\rho=(1-\phi)m_a/v_a+\phi m_b/v_b$.
Here $\phi=n_b v_b$ is the volume fraction of the solute. Incompressibility
implies that $n_a v_a+n_b v_b=1$.
Mass conservation can be expressed as
\begin{equation}
\frac{\partial \rho}{\partial t}+\nabla \cdot (\rho {\bf v})=0
\end{equation}
where the hydrodynamic flow velocity ${\bf v}$ is the velocity of the center of mass of local
volume elements. The particle currents can be decomposed in a center of mass
flux and a relative flux ${\bf j}={\bf j_b}=-{\bf j_a}$:
\begin{equation}
{\bf J}_i=n_i {\bf v}+\frac{{\bf j}_i}{m_i} \quad .
\end{equation}
Momentum conservation is described by the balance equation for the momentum
density $\rho {\bf v}$
\begin{equation}
\partial_t (\rho v_\alpha)-\partial_\beta \sigma_{\alpha\beta}=-\rho g \delta_{\alpha z}
\label{momentum}
\end{equation}
where the stress tensor $\sigma_{\alpha\beta}$ is the (negative) momentum flux tensor
and the momentum source corresponds to gravitational forces with gravitational acceleration $g$.
The z axis is oriented along the vertical direction,
opposite to the gravitational field.

Dissipation in the system is related to entropy production. The balance of the
entropy density $s$ reads
\begin{equation}
\frac{\partial s}{\partial t}+\nabla \cdot {\bf J}_s =  \theta
\end{equation}
where ${\bf J}_s$ is the entropy flux and $ \theta\geq 0$ is the local rate of entropy production
per unit volume. Since energy is conserved, the energy density $u$ obeys
\begin{equation}
\frac{\partial u}{\partial t}+\nabla \cdot {\bf J}_u = 0
\end{equation}
where ${\bf J}_u$ denotes the energy flux. 
The balance of the free energy density $f=u-Ts$ therefore reads
\begin{equation}
\frac{\partial f}{\partial t}+\nabla \cdot ({\bf J}_u-T{\bf J}_s) = -T  \theta
\end{equation}
For an isothermal system (which we consider here) with free energy density 
$f(n_a,n_b,{\bf v},z)=(1/2)\rho {\bf v}^2+
\rho g z+ f_0(n_a,n_b)$, 
the total entropy production rate $\dot S=\int d^3 r \theta$ is given by
\begin{eqnarray}
T\dot S&=&-\int d^3r \left (\frac{\partial}{\partial t}[\frac{1}{2}(\rho {\bf v}^2)+\rho g z]+
\sum_i \frac{\partial n_i}{\partial t} \mu_i\right )  +\int_{\partial \Omega} dA \;{\bf n}\cdot(T {\bf J}_s
-{\bf J}_u)
\\
&=& \int d^3r (\sigma_{\alpha\beta}^d u_{\alpha\beta}
-{\bf j}\cdot \nabla\bar \mu) \nonumber
\end{eqnarray}
where $\mu_i=\partial f_0/\partial n_i$ 
are the chemical potentials of the components.
The conservation laws and the Gibbs-Duhem relation $dP = n_a d\mu_a+n_b d\mu_b$
have been used to obtain the last line.
Here, $\partial\Omega$ denotes the boundary surface of the volume and
${\bf n}$ a vector normal to the boundary pointing out of the volume.
The hydrostatic pressure is denoted $P$.

The conjugate thermodynamic fluxes and forces are thus the pairs $\sigma^d_{\alpha\beta}$,
$u_{\alpha\beta}$ and ${\bf j}$,$-\nabla \bar \mu$, 
where $u_{\alpha\beta}=(\partial_\alpha v_\beta+\partial_\beta v_\alpha)/2$ is the tensor
of velocity gradients and
the dissipative part of the stress tensor is given by
$\sigma_{\alpha\beta}^d=\sigma_{\alpha\beta}-P\delta_{\alpha\beta}-\rho v_\alpha v_\beta$.
The relevant chemical potential is the difference
$\bar \mu=\mu_b/m_b-\mu_a/m_a$. 
Expressions for $\mu_a$ and $\mu_b$
in a simple model are given in Appendix A.

To linear order, the dissipative fluxes $\sigma^d_{\alpha\beta}$, $j_\alpha$ depend on the thermodynamic 
forces $u_{\alpha\beta}$, $\partial_\alpha \bar \mu$ as
\begin{eqnarray}
{j}_\alpha&=&-\gamma\partial_\alpha\bar\mu \label{jalpha}\\
\sigma^d_{\alpha\beta}&=&2\eta (u_{\alpha\beta}-\frac{1}{3}u_{\gamma\gamma}\delta_{\alpha\beta})
+\bar \eta u_{\gamma\gamma}\delta_{\alpha\beta} \label{sigmau}
\end{eqnarray}
Here, the viscosities $\eta$, $\bar \eta$  and the dissipative coefficient $\gamma$ have been
introduced.

{\bf Hydrodynamic equations}. Using Eq. (\ref{jalpha}),
the solute current can be expressed as
\begin{equation}
{\bf J}_b=-D\nabla n_b -\tilde\gamma \nabla P+ {\bf v}n_b \label{diffdr}
\end{equation}
Here, $D=(\gamma/m_b) \partial \bar \mu/\partial n_b\vert_P$ is the diffusion coefficient of the
solute and $\tilde\gamma =(\gamma/m_b) \partial\bar \mu/\partial P\vert_{n_b}\simeq
(\gamma/m_b)(v_a/m_a-v_b/m_b)$ describes the effects of
pressure gradients on solute molecules. In the limit of small solute concentration $n_b v_b\ll 1$, 
the dissipative coefficient scales as
$\gamma\simeq \mu m_b^2 n_b$, where $\mu$ is a mobility of solvent molecules.
This scaling implies that solvent molecules contribute independently to dissipation, no
interactions occur. In this limit, Eq. (\ref{diffdr}) becomes a drift-diffusion current
for independent solute molecules, see Appendix B.

From the force balance (\ref{momentum}) and 
the relation (\ref{sigmau}) follows the 
hydrodynamic equation of the 
barycentric
flow field
\begin{equation}
\eta \Delta  {\bf v} +(\bar\eta-\frac{\eta}{3})\nabla (\nabla\cdot {\bf v})  =-\nabla P + \rho g {\hat {\bf e}}_z
\label{eq:stokes}
\end{equation}
where ${\hat {\bf e}}_z$ is a unit vector in $z$-direction and inertial forces have been
neglected.
Incompressibility
of the fluid implies that the molecular volumes $v_a$ and $v_b$ are constant parameters. Since the density 
$\rho=(1-\phi)m_a/v_a+\phi m_b/v_b$
depends on solute volume fraction $\phi$, the divergence
\begin{equation}
 \nabla\cdot {\bf v}=-\frac{\Delta\rho}{\rho}\left ( \frac{\partial}{\partial t}+{\bf v}\cdot\nabla \right )\phi  \label{incomp}
\end{equation}
does not vanish. Here,
$\Delta\rho=m_b/v_b-m_a/v_a$ is the density difference of pure solute and
solvent.
In the incompressible system, 
the pressure profile $P$ plays the role of a Lagrange multiplier function
that is determined such that the corresponding flow satisfies the
incompressibility condition (\ref{incomp}).

{\bf Thermodynamic equilibrium}. 

The two-component fluid settles and eventually reaches an equilibrium state.
At equilibrium, both the flow velocity ${\bf v}$ and current 
${\bf J}_b$ vanish.  The hydrostatic pressure obeys
$\partial_zP=-g\rho(z)$. The solute  height profile $n_b(z)$ at equilibrium
satisfies the relation
\begin{equation}
\bar \mu (n_b(z),P(z))={\rm const.}
\end{equation}
and is independent on the dissipative coefficients.
In the limit of small solute concentration $n_b v_b\ll 1$, the solute height profile becomes a
barometric distribution $n_b(z)=\bar n e^{-z/\ell}$ with
characteristic length $\ell= k_B T/(\Delta\rho g v_b)$,
see Appendix B.

\section{Steady state gradients}

We now consider a simple geometry to discuss the main features
of steady flows and concentration gradients which can be 
maintained stationary over long times
in a two-component fluid but are intrinsically nonequilibrium states. 
A fluid filled channel of height $e$ along the $x$-axis, and infinite extension in $y$
direction,  is connected at both ends to reservoirs with different
solute concentrations $n_b$ or volume fractions $\phi=n_b v_b$, 
see Fig 1. 
The left reservoir is filled with fluid of solute volume fraction
$\phi_1$ up to a height $h_1$, the right reservoir up to a height
$h_2$ at volume fraction $\phi_2$. 
We consider well-stirred reservoirs such that the solute volume fraction
$\phi$ is uniform in each reservoir. Under such circumstances, $\phi$ is
also constant in time for short enough times as discussed below.

For these conditions, the mass densities in the reservoirs are
$\rho_{1,2}=(1-\phi_{1,2})m_a/v_a+\phi_{1,2}m_b/v_b$.
The pressures in the reservoirs are
given by $P_{1}=\rho_{1} g (h_{1} -z)+P_0$ and
$P_{2}=\rho_{2} g (h_{2} -z)+P_0$, where $P_0$ denotes the outside pressure.
\begin{figure}[htbp]
\includegraphics[width=6cm]{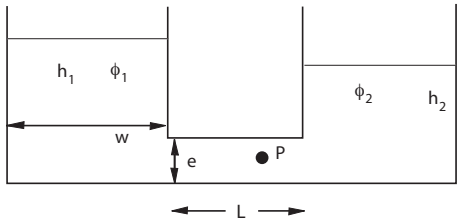}
\caption{Fluid channel of height $e$ and length $L$
between two reservoirs of width $w$, filled with fluid of volume fractions
$\phi_1$ and $\phi_2$ and heights $h_1$ and $h_2$. 
In such a setting, 
gradients of pressure or concentration can be maintained over long times.
Such gradients can drive the 
motion of an immersed colloidal particle $P$.}
\label{fig}
\end{figure}
For $h_{1,2}\gg e$,  the pressure gradient in the channel generated by the reservoirs
is given by
\begin{equation}
\partial_x P\simeq g(\rho_2 h_2-\rho_1 h_1)/L
\end{equation}
We consider in the following two complementary cases. 

{\bf Case (A): concentration gradient in the absence of barycentric flows}.
In this case, the pressures on each side are balanced,
$h_1\rho_1=h_2\rho_2$, and therefore, $\partial_x P=0$, and
no flow exists, ${\bf v}=0$. In this case,
after an initial relaxation process of duration $t_r\simeq L^2/D$,
the concentration profile becomes a linear gradient
\begin{equation}
\partial_x \phi= (\phi_2-\phi_1)/L\quad .
\end{equation}
The corresponding diffusion current is
\begin{equation}
(J_b)_x=-D \frac{\phi_2-\phi_1}{Lv_b}
\end{equation}
This current is maintained on time scales short compared to the
equilibration time between reservoirs $t_d\simeq hwL/(eD)$, 
where $w$ and $h$ are reservoir width and height respectively. This
time can be made arbitrarily long.

 
{\bf Case (B): flow without concentration gradient.} In this case, 
the volume fractions and mass densities are equal in both reservoirs, $\phi_1=\phi_2$ and
$\rho=\rho_1=\rho_2$. For $h_1\neq h_2$, a
pressure gradient and a corresponding hydrodynamic flow exists  ${\bf v}=(v_x,0,0)$
with $\nabla\cdot {\bf v}=0$.
The corresponding solution to Eq (\ref{eq:stokes}) is
\begin{equation}
 v_x(z)\simeq \frac{1}{2}\frac{g\rho( h_2-h_1)}{\eta L} z(z-e) \label{flow}
\end{equation}
In this situation, the solute concentration $n_b$ in the channel is constant.  
A solute current exists,
$({ J}_b)_x=n_b v_x(z)+\tilde \gamma \partial_x P$,  which consists of a convective part and a dissipative flux relative
to the fluid flow.
This barycentric flow and solute current are maintained during the
shortest of the times $t_d\simeq hwL/(eD)$ and $t_c\simeq wL\eta/(e^3 \rho g)$.
Note again that these times can be arbitrarily large.
This linear concentration gradient also applies to good approximation for weak
flows if $D/v_x\gg L$. For larger flow velocities, nonlinear concentration
fields occur.

\section{Dissipative interfacial processes: surface slip velocities}

A colloidal particle that is immersed in a two component fluid as described above
can be set in motion as a result of nonequilibrium conditions provided
by either a pressure gradient or a solute gradient. In order to determine
the velocity of motion of the bead relative to the flow, and the corresponding 
perturbation of the flow field,
the hydrodynamic equations of the fluid are solved with appropriate
boundary conditions applied at the particle surface. These boundary conditions
can be systematically derived by expressing surface dissipation
at the fluid particle interface and writing generic Onsager relations for
conjugate forces and fluxes at the interface.

We consider a solid substrate in contact with a two-component
fluid. We use a coordinate system with $z$ denoting the distance from 
the solid surface along the normal direction. In the interface, 
we average all quantities over the
thickness $d$ in which properties differ from bulk properties. This
procedure is valid when $d\ll R$, where $R^{-1}$ is a local curvature of the
interface.
The interface dissipation rate reads (see Appendix D)
\begin{equation}
T \dot S=\int dA (\frac{1}{2}\sigma^s_{iz}v^s_i-\;   j^s_i \nabla_i \bar\mu^s + r^s \Delta \mu)
\end{equation}
Here,  the indicies $i,j$ denote directions parallel to the interface. The superscripts $s$ indicate that surface fields
are considered,  $v_i^s=v_i(z=0)$ and $\sigma_{iz}^s=\sigma_{iz}(z=0)$. 
Note that we have neglected for simplicity surface viscosity
and $j^s_i$ is a surface current with
units of mass per length and time. To keep our discussion simple, we consider
here the case where the interface equilibrates rapidly with the fluid \cite{ande89},
$\bar \mu^s = \bar \mu(z=0)$ and material exchange between interface and fluid
can be negected $v_z(z=0)=0$ and $j_z(z=0)=0$.

In addition to the two conjugate fluxes and forces already discussed,
we have added here the term $r^s\Delta\mu$ which describes 
an 
active process on the interface which can propel
a swimmer. An example from biology would be a large number beating of cilia on the particle 
surface.  The active process is driven by a chemical fuel (which in a living cell would be
ATP) with chemical potential difference between fuel and product $\Delta \mu$.
The conjugate flux $r^s$ denotes the rate of fuel consumption per unit area of the surface. 
The active process
can only contribute to net motion generation if the surface has a vectorial asymmetry
which for example determines the direction of the active stroke of a cilium along the surface. 
This direction is characterized by a normalized vector
$p_i$  tangent to the surface.

The
three pairs of conjugate fluxes and forces are related to linear order by
\begin{eqnarray}
v^s_{i}&=&\nu \sigma^s_{iz} -\alpha \nabla_i\bar\mu^s + p_i \zeta \Delta \mu \label{eq:nu}\\
 j^s_i &=&-\alpha \sigma_{iz}-\gamma^s \nabla_i \bar\mu^s +p_i \zeta' \Delta\mu \\
 r^s &=& \zeta p_i \sigma^s_{iz}+ \zeta' p_i \nabla_i \bar \mu^s +  \Lambda \Delta \mu \quad ,
\end{eqnarray}
where the coefficients satisfy Onsager symmetry relations. 
We observe that there exists in general a finite slip velocity $v_i^s$ at the surface. 
The coefficient
$\nu$ can be characterized by the related "slip length"  $b=\nu\eta$, which is the 
distance from the surface at which
an effective no slip boundary condition applies.
The coefficient $\gamma^s$ is related to surface diffusion and
$\alpha$ is a dissipative coefficient which couples surface flows to
relative fluxes between solvent and solute. The coefficients $\zeta$ and
$\zeta'$ describe the coupling of the active process on the surface to the
two-component fluid.

We now consider the case $\zeta=0$ and $\zeta'=0$.
We furthermore assume that the slip length vanishes, $\nu=0$, which implies that there
is no slip in the absence of a chemical potential gradient. There remains a
surface slip velocity driven by chemical potential gradients and pressure gradients,
which as in the bulk can be expressed as 
\begin{equation}
v_i^s=-\kappa \nabla_i n_b^s - \kappa' \nabla_i P^s \label{slip}
\end{equation}
Here $\kappa=(\alpha/m_b) (\partial \bar \mu/\partial n_b)\vert_{z=0}$ and 
$\kappa'=(\alpha/m_b) (\partial \bar \mu/\partial P)\vert_{z=0}$.
For a specific interface model, the coefficient $\kappa$ has been
expressed as $\kappa\simeq k_B T \lambda^2/\eta$ \cite{ande89}, where the length scale $\lambda$ 
is related to the range of the potential describing interactions 
between solute molecules and the particle
surface and is of the order of the interface thickness $d$.


\section{Colloidal transport}

We now consider a spherical colloidal particle of radius $a$ which is
subject to the solute concentration gradient or pressure gradient. 
We are interested in the 
particle velocity $v_p=v_x+\Delta v$ in $x$-direction
in the laboratory frame where $\Delta v$ is the velocity difference between
particle and flow. These velocities 
can be determined
by solving the Stokes equation (\ref{eq:stokes}) 
with the appropriate slip
boundary conditions on the particle surface.

In the following, we consider the case where the particle is far from
any walls which allows us to ignore the effects of boundaries
of the channel in which the particle is placed.
We ignore sedimentation of the particle in the gravitational field
in $z$-direction. 
We discuss two cases (A) and (B) described
above. 

{\bf (A) Concentration gradient in the absence of barycentric flows}.  
In the presence of a constant concentration gradient
$\partial_x n_b=(\phi_1-\phi_2)/(Lv_b)$,  a 
slip velocity at the particle surface is generated.
We consider the case of zero slip length $b=0$ and no
active process $\Delta\mu=0$. 
In spherical coordinates of the particle $r,\theta$, where $\theta$ measures
the angle with respect to the gradient direction along the $x$-axis, and $r$ is the
radial distance from the particle center, the slip velocity ${\bf v}^s=v_s(\theta) {\bf e}_{\theta}$,
is given by
\begin{equation}
v_s(\theta)=-\kappa \sin(\theta) \partial_{x} n_b 
\end{equation}
where ${\bf e}_\theta$ is a unit vector in $\theta$ direction tangential to the sphere.
Using the boundary conditions $v_\theta(r=a,\theta)=v_s(\theta)$ and
$v_r(r=a,\theta)=0$,
the flow field around a spherical particle is given by (see Appendix E)
\begin{eqnarray}
v_r(r,\theta)&=& - \Delta v \left (1-\frac{a^3}{r^3}\right )\cos(\theta) \nonumber \\
v_{\theta}(r,\theta)&=&\frac{\Delta v}{2} \left (2+\frac{a^3}{r^3}\right ) \sin(\theta)\label{swimflow}
\end{eqnarray}
where we have used for simplicity the condition $\nabla\cdot {\bf v}=0$ which is
according to Eq. (\ref{incomp}) satisfied if $v_a/m_a=v_b/m_b$.
The relative velocity between particle and fluid far from the particle is
\begin{equation}
\Delta v = (2\kappa/3) \partial_x n_b \quad .
\end{equation}
There is no body force acting on the particle,  $f^p=0$, see Appendices E and F. %
The flow field perturbation decays as $\sim 1/r^3$ for increasing $r$ and thus faster than
a Stokeslet which decays as $\sim 1/r$  and is the signature of a net force.
The decay $\sim 1/r^3$ corresponds to a source doublet and  
implies that a force quadrupole is exerted by the particle on the fluid.
Here, no force dipole exists.
In more general situations, a force dipole also exists,
which correspond to a decay
of the velocity $\sim 1/r^2$, see Appendix E. Note that a force dipole
does not contribute to propulsion but dominates the far field.
%
Note also, 
that there are no forces due to osmotic pressures acting on a particle in a concentration gradient,
see Appendix C. 

{\bf (B) Flows in the absence of concentration gradients:} 
In the presence of a pressure gradient $\partial_x P$, a parabolic flow
profile (\ref{flow}) is generated. If we can ignore the effects of walls, 
and if no surface slip occurs,
the
relative velocity of the particle with respect to the flow $\Delta v_1=v_p-v_x$ can be determined
by Faxens law as \cite{happel} (see Appendix E)
\begin{equation}
\Delta v_1 = \frac{3a^2\partial_x P}{4\eta} \quad .
\end{equation}
Surface slip induces an additional component $\Delta v_2$ to the relative motion between
fluid and particle.
For vanishing slip length $b=0$, the only contribution to the slip
is $v^s_i=-\kappa'\nabla_i P^s$. The boundary conditions in a coordinate frame co-moving
with the particle are $v_{r}(\theta, r=a)=0$ and
$v_{\theta}(\theta, r=a)=-\kappa' \sin(\theta)\partial_x P$. The flow field perturbation generated by
surface slip is given by Eq. (\ref{swimflow}) with $\Delta v=\Delta v_2=(2\kappa'/3)\partial_x P$.

Since $\kappa'$ expresses the momentum transfer to the particle
within the interfacial layer of thickness $d$, one expects $\kappa'\simeq -\bar d^2/\eta$,
where the length $\bar d$ is of order $d$.
By superimposing the flow field (\ref{flow}) with the perturbation 
by the particle $\Delta v_1$ and $\Delta v_2$, the overall particle velocity is
\begin{equation}
v_p\simeq -\left [\frac{e^2}{8\eta}-\frac{a^2 }{12\eta}+\frac{2 \bar d^2}{3\eta}\right] 
\frac{g(\rho_2 h_2-\rho_1 h_1)}{L}
\end{equation}
which is entirely driven by the pressure gradient. No net body force $f^p$ acts on the particle,
see Appendix F. Note that the convective
term is large compared to $\Delta v_1$ which is in turn large compared to $\Delta v_2$.
The existence of $\Delta v_1$ is yet important since it can lead to
separation of particles according to their size. 

\section{Self-propulsion}

{\bf Propulsion by self-generated concentration gradients}. A chemical reaction catalyzed on the
particle surface can generate local concentration gradients which propel the
particle \cite{gole05,gole07}. To describe situations where the particle catalyzes a reaction involving
two solvent species, at least a three-component fluid description is required.
However, the basic physics of self-propulsion can be captured by our two-component
fluid  if we assume that
a surface reaction transforms molecules of type $a$ into molecules of type $b$.
In this case, we consider two components with equal molecular masses $m_a=m_b$ to satisfy mass conservation.
The molecular volumes $v_a$ and $v_b$ can in general differ. 
A local reaction rate $S$ per unit
area at which $a$ molecules are transformed in $b$ molecules on the particle surface
implies the boundary 
conditions for the molecular fluxes normal to the particle surface
\begin{equation}
({J_b})_z(z=0)=-(J_a)_z(z=0)=S
\end{equation}
in the reference frame where the particle is at rest.
The corresponding boundary condition for the center of
mass velocity is $v_z(z=0)=0$.

The nonlinear convective term in the flux Eq. (\ref{diffdr}), couples the concentration
and flow fields. In the limit of small Peclet number ${\rm Pe}=\Delta v \;a/D$, 
we can neglect this convective nonlinearity. 
The stationary concentration
field is then solution to the diffusion equation $\nabla^2 \phi=0$ \cite{gole05,gole07}.  
For the simple choice of an asymmetrically
distributed reaction rate $S(\theta)=S_0 \cos(\theta)$, 
the solution for the boundary conditions specified above is given by
\begin{equation}
n_b(r,\theta)=\frac{S_0 a^3}{2 D  r^2} \cos(\theta)+n_b^\infty
\end{equation}
where $n_b^{\infty}$ denotes the concentration far from the particle.
This concentration field induces according to Eq. (\ref{slip}) the surface slip velocity 
\begin{equation}
v_s(\theta)=\frac{\kappa S_0}{2D}\sin(\theta) 
\end{equation}
which is independent of particle radius. The corresponding hydrodynamic flow field
for the case $v_a=v_b$ for which $\nabla \cdot {\bf v}=0$ is given by Eq. (\ref{swimflow}) with
\begin{equation}
\Delta v=\frac{\kappa S_0}{3D}\quad .\label{Dv}
\end{equation}
The pressure $P$ is constant and there is no net body force $f^p$ acting on the
particle. 


For finite Peclet number, there exist no 
simple solutions to this nonlinear problem since the convection velocity in Eq. (\ref{diffdr}) 
is given by the hydrodynamic flow field described by Eq. (\ref{eq:stokes}). The latter in turn is coupled
to the concentration field $n_b$ via the surface slip driven by the
chemical reaction. Here, 
we focus on the scaling behavior of the propulsion velocity 
for large ${\rm Pe}$. 
The hydrodynamic flow
relative to the particle is of order $\Delta v$.
This defines a length scale $\ell \simeq( Da/\Delta v)^{1/2}\ll a$ which characterizes the thickness
of the boundary layer near the particle in which the concentration field is different from $n_b^\infty$
by a variation of order $\Delta n_b \simeq \ell S_0/D$. The resulting slip velocity
is of the order $\Delta v\simeq \kappa \Delta n_b/a$. By combining these expressions, we obtain
\begin{equation}
\Delta v\simeq \frac{\kappa^{2/3} S_0^{2/3}}{D^{1/3}a^{1/3}} \label{Dvsc}
\end{equation} 
The velocity thus increases for large ${\rm Pe}$ less than linearly and becomes
dependent on the particle size $a$. For ${\rm Pe}=1$, the relations (\ref{Dv})
and (\ref{Dvsc}) match except for a dimensionless prefactor.

{\bf Propulsion by active surface processes}.
Many microorganisms swim by using a large number of cilia attached to the surface
which generate periodic beating movements that are driven by molecular motors that
consume a chemical fuel \cite{bray92}. 
In a coarse-grained picture where the cilia are active elements within the interface between swimmer and fluid, 
the motion of the cilia effectively
generates a surface slip velocity of the flow. In our generic description
this surface slip is captured by the term $v_i^s\simeq p_i \zeta\Delta \mu$
in Eq. (\ref{eq:nu}). Here, the tangent vector $p_i$ describes the direction along which
the cilium generates a flow and $\zeta$ is a coupling coefficient between the
free energy $\Delta\mu$ driving the motors and the surface flow. 
In this scenario, the slip pattern on the surface of
the swimmer is determined by the ciliar beat direction and strength. 

For given surface slip pattern and corresponding hydrodynamic flows with $\nabla\cdot {\bf v}=0$, 
the velocity of the swimmer can be determined without explicit calculation of the flow field. For
a spherical particle it is given by \cite{ston96}
\begin{equation}
\Delta {\bf v} = \frac{1}{4\pi a^2}\int dA \;{\bf v}^s
\end{equation}
Similarly, the rotation rate and axis (which can exist for
a surface slip velocity field lacking axial symmetry) is described  
by \cite{ston96}
\begin{equation}
{\bf \Omega} = \frac{3}{8\pi a^3}\int dA\; {\bf n}\times{\bf v}^s
\end{equation}
where ${\bf n}$ is a unit vector normal to the surface and ${\bf \Omega}$ points in the
direction of the rotation axis.
Again, this translational and rotational motion occurs relative to the fluid without a net body force and torque acting
on the swimmer.
For the simple example with 
$p_i$ a unit tangent vector in $\theta$ direction
and $\zeta=\zeta_0\sin(\theta)$, we have $\Delta v=(2/3) \zeta_0\Delta\mu$
and again the flow field perturbation of Eq. (\ref{swimflow}).

\section{Discussion}

We have presented a generic description of colloidal transport
to clarify the force balances involved and the role of interfacial slip. 
The simplest form of transport occurs if
relative external body forces $f^p$ are applied to a particle. External body forces, such as those
due to gravitation, correspond to
source terms in the momentum balance (\ref{momentum}).
As a result, the particle moves at a speed $\Delta v$ relative to the fluid and experiences
Stokes friction $f^p = 6\pi \eta a \Delta v$. The corresponding
perturbation of the fluid flow at large distances is given by a Stokeslet
which decays as $1/r$.

In the absence of external fields, a colloidal particle in a fluid is force free, $f^p=0$, 
even if concentration or pressure %
gradients exist. 
In this case a colloidal particle nevertheless moves relative to the
fluid at a velocity $\Delta v$ if there is a slip of the flow at the surface of the particle. Such 
slip is generated by surface dissipative phenomena 
\cite{ande89}. We have shown that 
Onsager relations on the solid surface determine the boundary
conditions for the hydrodynamic equations in the bulk. This description
can account for a variety of phenomena, including a slip length, slip
induced by concentration or pressure gradients as well as slip due to active
processes on the particle surface, within a unified framework.

Our generic description can be generalized to electric fields, which requires to
replace chemical potentials by electrochemical potentials and to
include electrostatics in the free energy density.
Effects in the presence of electric fields include electrophoresis of
charged particles. Since a
charged particle is screened beyond an electric double layer of counter ions,
it is effectively neutral.
Electrophoresis is thus another example where no force acts on a
particle (including the layer of counter ions) and electrophoretic motion results thus from surface
slip \cite{ande89,long98}.

Our arguments are relevant for mechanisms of self-propulsion
of colloidal particles. Self propulsion implies that motion occurs in the
absence of externally applied forces. 
This is possible if a particle self-generates a surface
slip velocity by active processes on or near the particle surface. These include
the action of cilia and flagella in the case of swimming microorganisms \cite{bray92}
or the generation of a concentration gradient by surface chemical
reactions \cite{hows07}.  The main effect of the surface slip 
is to generate a relative motion between particle and fluid. 
In the general case, this relative motion 
is associated with a flow field perturbation that
decays as $\sim 1/r^2$, corresponding to a force dipole exerted by
the particle on the fluid which however does not contribute to propulsion. 
If no force dipole exists, such as in the case described
by Eq. (\ref{swimflow}), the flow perturbation decays as $\sim 1/r^3$. This
flow perturbation  corresponds to a force quadrupole acting on the fluid.
In all cases there is
no net external force acting on the particle. 

In a recent publication it has been claimed that self-generated concentration
gradients generate a force on a particle by osmotic pressure gradients 
which is balanced by Stokes friction \cite{cord08}.
Clearly, no force is exerted on a particle in a concentration
field by osmotic pressure, see Appendix C.
The scenario proposed in \cite{cord08}
violates momentum conservation, see Appendix F.  
The work of Ref. \cite{cord08} furthermore 
ignores the effect of hydrodynamic
flow perturbations  on the concentration gradients, 
when those cannot be neglected at finite Peclet numbers.
Our generic description of colloidal transport can serve to clarify
these and related points in a systematic way.

We thank Ramin Golestanian for stimulating discussions.
F.J. thanks Friederike Schmidt, Holger Stark and Andrej Vilfan for similarly stimulating 
discussions on the hydrodynamic flow perturbations generated by colloidal particles and 
swimmers. 

\appendix
\section{Entropy of mixing and chemical potentials}

The properties of the chemical potentials can be discussed using a
simple model for a two-component fluid with a free energy in the rest
frame
$F_0(N_a,N_b,V)=V f_0(N_a/V,N_b/V)$, where $V$ denotes volume, with
\begin{equation}
f_0(n_a,n_b)=k_B T(n_a \ln\frac{n_a v_a}{n_a v_a+n_b v_b}
+n_b \ln\frac{n_b v_b}{n_a v_a+n_b v_b})
+\frac{\chi}{2}(n_a v_a+n_b v_b-1)^2
\end{equation}
Here, the first term describes the entropy of mixing of two
components with molecular volumes $v_a$ and $v_b$, the
second term describes the compressibility of the fluid
by the coefficient $\chi$. Interactions between the two components are
neglected. The chemical potentials $\mu_i=\partial f_0/\partial n_i$
are given by
\begin{eqnarray}
\mu_a(n_a,n_b)&= &k_B T \left( \ln\frac{n_av_a}{n_av_a+n_bv_b}
+\frac{n_b(v_b-v_a)}{n_av_a+n_bv_b}\right )+v_a \chi(n_av_a+n_bv_b-1) \label{mua0}\\
\mu_b(n_a,n_b)&= &k_B T \left ( \ln\frac{n_bv_b}{n_av_a+n_bv_b}
+\frac{n_a(v_a-v_b)}{n_av_a+n_bv_b}\right )+v_b \chi(n_av_a+n_bv_b-1)
\end{eqnarray}
The pressure $P=-(\partial F_0/\partial V)\vert_{N_a,N_b}=-f_0+\mu_a n_a+\mu_b n_b$ is
\begin{equation}
P(n_a,n_b)=\frac{\chi}{2}((n_av_a+n_bv_b)^2-1)
\end{equation}
Using the pressure, the solvent density $n_a$ can be eliminated. In the incompressible
limit of large $\chi$, $n_a v_a+n_b v_b=1$ and we obtain
\begin{eqnarray}
\mu_a(n_b,P)&\simeq &k_B T ( \ln(1-n_bv_b)+n_b(v_b-v_a))+P v_a \label{mua}\\
\mu_b(n_b,P)&\simeq &k_B T ( \ln(n_bv_b)+n_a(v_a-v_b))+P v_b
\end{eqnarray}

\section{Limit of small solute concentration}

The solute flux is driven by gradients of the chemical potential difference
$\bar \mu=\mu_b/m_b-\mu_a/m_a$;
\begin{equation}
{\bf J}_b=-\frac{\gamma}{m_b}\nabla \bar \mu+n_b{\bf v}
\end{equation}
In the limit $n_b v_b\ll 1$ of small solute concentration, 
\begin{equation}
\bar\mu(n_b,P)\simeq 
\frac{k_B T}{ m_b}\ln (n_b v_b) 
+\left (\frac{v_b}{m_b}-\frac{v_a}{m_a}\right )P 
\end{equation}
For small $n_b$, solute particles become independent of each other and dissipation
takes place independently for each solute particle. 
Therefore, $\gamma\simeq \xi m_b^2 n_b$, where $\xi$ is a
mobility per solute molecule.
We thus find
\begin{equation}
{\bf J}_b\simeq -D \nabla n_b - \bar \gamma n_b \nabla P +n_b{\bf v}
\end{equation} 
where 
\begin{eqnarray}
\bar \gamma&\simeq &\xi m_b (v_b/m_b-v_a/m_a)\nonumber \\
D &\simeq &\xi k_B T \label{Dgamma}
\end{eqnarray}
In the limit of small $n_bv_b$, the pressure gradient is approximately constant
$\partial_z P=-\rho g \simeq -g m_a/v_a$. For ${\bf v}=0$, the height profile
is
\begin{equation}
n_b=\bar n e^{-z/\ell} \label{nz}
\end{equation} 
with $\ell=D v_a/(\bar\gamma m_a g)$. 
Using Eq. (\ref{Dgamma}), we obtain $\ell=k_B T/(\Delta\rho g v_b)$
and (\ref{nz}) is the barometric height distribution.

\section{Osmotic pressure}

Osmotic pressures are a consequence of a semipermeable interface which separates
the fluid in two compartments $(1)$ and $(2)$. The solvent
passes this interface, which is impermeable to the solute. As a consequence,
across the interface the chemical potential of the solvent is balanced, 
$\mu_a^{(1)}=\mu_a^{(2)}$.
However, the solute chemical potentials do not balance 
$\mu_b^{(1)}\neq \mu_b^{(2)}$.

The solvent chemical potential is according to Eq. (\ref{mua}) in the limit of small
$n_b v_b $ 
given by $\mu_a\simeq -k_B T n_bv_a
+P v_a$. The balance of solvent chemical potentials implies
the existence of an osmotic pressure difference across the semipermeable membrane
\begin{equation}
P^{(2)}-P^{(1)}=k_B T (n_b^{(2)}-n_b^{(1)}) \quad .
\end{equation}
Note that the hydrostatic pressure difference appears only after the balance of the chemical potential of the solvent is
reached and that the corresponding momentum source is provided by the membrane.
\section{Force balance and dissipation at an interface}

We consider dissipation and force balances in an interfacial region
of thickness $d$ between two phases in which material properties differ from
those in the two bulk phases. 
The coordinate normal to the interface
is denoted $z$.
A relative slip velocity can occur at an interface. 
The local center of mass velocity tangential to the interface at $z=\pm d/2$ 
is denoted $v_i^{\pm}$, respectively.
Dissipation due to interfacial slip
\begin{equation}
v^s_i=v_i^+-v_i^-
\end{equation}
can be expressed as
\begin{equation}
T\dot S \simeq  \int dA \int_{-d/2}^{d/2}dz\; \frac{\partial_z v_i}{2}\; \sigma_{iz} \simeq \frac{1}{2}\int dA  \;\sigma_{iz}^s v_i^s
\end{equation}

The interfacial shear stress $\sigma_{iz}^s=\epsilon \sigma^{+}_{iz}+(1-\epsilon)\sigma^{-}_{iz}$ is a weighted average of $\sigma_{iz}^\pm$. The value of $0<\epsilon<1$ depends on the internal structure of
the interface. 
The conjugate thermodynamic variables are
thus $v_i^s$ and $\sigma^s_{iz}$. The corresponding Onsager relation reads
\begin{equation}
v_i^+-v_i^- = \nu(\epsilon \sigma^+_{iz}+(1-\epsilon)\sigma_{iz}^-)
\end{equation}
where $\nu$ is the corresponding dissipative coefficient. 

The force balance in the interfacial region $\partial_\beta \sigma_{\alpha\beta}=0$
implies
\begin{equation}
\int _{-d/2}^{d/2}dz\;( \partial_z \sigma_{iz} + \partial_j \sigma_{ij})=0
\end{equation}
This implies the interfacial force balance
\begin{equation}
\sigma_{iz}^+-\sigma_{iz}^-=-\partial_i \Sigma
\end{equation}
where the interfacial tension (for isotropic stresses in the tangent plane) is
\begin{equation}
\Sigma=\frac{1}{2}\int_{-d/2}^{d/2} dz\; \sigma_{kk}
\end{equation}
In the absence of interfacial tension gradients $\partial_i\Sigma$, the
shear stress is continuous across the interface $\sigma^s_{iz}=\sigma_{iz}^+=\sigma_{iz}^-$.
and the slip velocity is simply given by $v_i^s=\nu \sigma^+_{iz}$ as in Eq. (\ref{eq:nu}).

\section{Hydrodynamic flow fields with axial symmetry}

Solutions to the Stokes Eq. (\ref{eq:stokes}) for axisymmetric 
incompressible flows with
$\nabla\cdot {\bf v}=0$ can be
expressed using the stream function $\psi$ \cite{happel}. 
In spherical coordinates,
the velocity field is related to the stream
function  $\psi(r,\theta)$ by
\begin{eqnarray}
v_r&=&-\frac{1}{r^2 \sin\theta}\frac{\partial \psi}{\partial \theta} \\
v_{\theta}&=&\frac{1}{r \sin\theta}\frac{\partial \psi}{\partial r}
\end{eqnarray}
The stream function satisfies the differential equation
$E^4 \psi=0$, where
\begin{equation}
E^2 \psi=\left ( \frac{\partial^2}{\partial r^2}+\frac{\sin\theta}{r^2}\frac{\partial}{\partial \theta}
\frac{1}{\sin\theta}\frac{\partial}{\partial \theta} \right ) \psi
\end{equation}
Simple solutions are given by
\begin{equation}
\psi=\sin^2 \theta (A_1 r^4+A_2 r^2 +A_3 r + \frac{A_4}{r}) \label{eq:flow}
\end{equation}
where $A_1,\dots, A_4$ are constant parameters determined
by boundary conditions. 
The corresponding pressure field is given by $P=- \eta \cos\theta (20 A_1 r+2A_3 /r^2)+P^{\infty}$,
where $P^{\infty}$ is the pressure far from the particle. The body force acting on the particle
which 
is balanced by forces exerted by the hydrodynamic flow
is $f^p=-8\pi \eta A_3$ \cite{happel}.

In case (A), we determine
a solution of the form given by Eq. (\ref{eq:flow}) with $v_\theta (r=a)=v_s(\theta)$
and $v_r(r=a)=0$ in the reference frame moving with the sphere. 
For large $r$, we require motion at constant
velocity $\Delta v$ in negative $x$-direction, $v_r \simeq -\Delta v \cos(\theta)$
and $v_\theta\simeq \Delta v \sin\theta$. From the latter conditions,
it follows that $A_1=0$, and $2A_2=-\Delta v$. 
Because there is no external force acting on the particle, $f^p=0$ and thus $A_3=0$. 
The boundary conditions on the
particle surface imply $A_4=-A_2a^3$ and $3A_2 =-(\kappa/v_b)\partial_x\phi$.
The corresponding
flow is given by Eq. (\ref{swimflow}). The perturbation of the flow velocity decaying as $\sim 1/r^3$ 
corresponds to a source doublet \cite{happel,blak71} and implies that a force quadrupole is exerted by the
particle on the fluid.

In case (B),  we superimpose the flow (\ref{flow}) driven by an applied 
pressure gradient with the velocity $\Delta v_1$ due to the perturbation of
the Poiseuille flow by the sphere and the relative velocity $\Delta v_2$
due to interfacial slip. 
The velocity $\Delta v_1$ can be estimated from Faxens theorem 
which expresses the force on the particle as
\begin{equation}
{f}^p_x=6 \pi \eta a (v_p-v^0_x)+\pi a^3 \nabla^2 {v^0_x}
\end{equation}
where $v^0_x$ denotes the unperturbed parabolic flow field.
No force acts on the particle, $f^p_x=0$, which determines the relative
velocity $\Delta v_1=v_p-v_x^0$. 
This flow is superimposed with  a flow 
driven by slip boundary conditions (\ref{swimflow}) as described for case (A).

If a net body force $f^p$ acts on the particle with no slip, 
the boundary conditions are $v_r(r=a)=0$ and $v_\theta(r=a)=0$ which require
$3A_4=A_3a^2$ and $A_2=-(2/3)A_3$.
From the asymptotic behavior, it follows that $A_1=0$ and $2A_2=-\Delta v$. 
This implies $f^p=6\pi \eta a \Delta v$ and the flow field is given by
\begin{eqnarray}
v_r(r,\theta)&=& - \Delta v \left (1-\frac{3}{2}\frac{a}{r} +\frac{a^3}{2r^3} \right )\cos(\theta) \nonumber \\
v_{\theta}(r,\theta)&=& \Delta v \left (1-\frac{3}{4}\frac{a}{r}-\frac{a^3}{4r^3}\right ) \sin(\theta)\label{stokesflow}
\end{eqnarray}
In addition to the force monopole this flow contains again a contribution from a source doublet
implying a force quadrupole.

The equation $E^4 \psi=0$ for the flow also has the solution 
\begin{equation}
\psi = -\cos\theta\sin^2\theta (B_1+B_2/r^2) \quad .\label{fd}
\end{equation}
The corresponding flow field is $v_r=(3\cos^2\theta-1)(B_1/r^2+B_2/r^4)$ and 
$v_\theta=B_2 \sin(2\theta) /(2 r^4)$ \cite{blak71a}. The condition $v_r(r=a)=0$
that the radial flow vanishes on the particle surface imposes $B_2=-B_1a^2$.
The decay
of the radial component proportional to $\sim 1/r^2$ corresponds to a stokes doublet
which implies the action of
a force dipole on the fluid \cite{blak71}. This force dipole 
dominates in the far field over the force quadarupole. 
For an arbitrary distribution of surface slip $v_\theta(r=a) = v_s(\theta)$, a force dipole exists
in general. The force dipole vanishes by symmetry if the surface slip is a symmetric
function $v_s(\theta)=v_s(\pi-\theta)$ such as is the case for $v_s\sim \sin(\theta)$ 
described by  Eq. (\ref{swimflow}). Note that the flow perturbation corresponding to
a force dipole does not contribute to propulsion by symmetry. 

\section{General considerations concerning the existence or non-existence of a Stokeslet in a velocity field  
carrying a particle}

Consider a particle of any shape, including arbitrary topological genus, immersed in a multi-component fluid flowing in a container of complex geometry possibly of non trivial topology. The fluid is submitted to an external force 
of density $ g_\alpha^{f,{\rm ext}}$  and the particle to the force density $g_\alpha^{p,{\rm ext}}$. 
The momentum flux in the fluid is characterized by a stress tensor $\sigma_{\alpha\beta}$. The total force acting on the particle reads:
 \begin{equation}
 f_\alpha^{\rm tot} = \int_{V_p}  g_\alpha^{p,{\rm ext}}dV+\int_{S_p} \sigma_{\alpha\beta} dA_\beta
 \end{equation}
Here, 
the integration volume $V_p$ and surface $S_p$
refer to the particle and the surface elements $dA_\beta$ are oriented to point outward from the particle.
The particle dynamics is $m\ddot x_\alpha = f_\alpha^{\rm tot}$, where $x_\alpha$ is the particle
position and $m$ is the particle mass.
 In the Stokes limit where inertial terms can be neglected or in stationary conditions $f_\alpha^{\rm tot}=0$,
 which implies that in inertia free regimes the total force acting on a particle in a fluid vanishes.
 Momentum conservation in the fluid implies
 \begin{equation}
 g_\alpha^{f,{\rm ext}}+\partial_\beta \sigma_{\alpha\beta}=0
 \end{equation}
 where inertial forces have again been neglected.
 In integral form, this can be expressed as
\begin{equation}
\int_{S_p}\sigma_{\alpha\beta}dA_\beta+\int_S \sigma_{\alpha\beta}dA_\beta=
-\int_{V-V_p}g_\alpha^{f,{\rm ext}}dV
\end{equation}
Here the integration is over the particle surface $S_p$ and an arbitrary surface $S$
enclosing a volume $V$ which includes the particle in the fluid. The 
surface elements are oriented to point to the outside of the fluid
volume $V-V_p$ between the surfaces $S_p$ and $S$.
The (vanishing) total force now reads
\begin{equation}
f_\alpha^{\rm tot}=\int_{V_p}g_\alpha^{p,{\rm ext}}dV+\int_S\sigma_{\alpha\beta}dA_\beta+\int_{V-V_p}g_\alpha^{f,{\rm ext}}dV
\end{equation}  
or equivalently,
\begin{equation}
f_\alpha^{\rm tot}=\int_{V_p}(g_\alpha^{p,{\rm ext}}-g_\alpha^{f,{\rm ext}})dV+\int_S\sigma_{\alpha\beta}dA_\beta+\int_{V}g_\alpha^{f,{\rm ext}}dV
\end{equation}
Consider now the stress $\sigma_{ij}^0$ 
in the fluid in the absence of the particle (every other condition being kept identical). 
For any volume and corresponding surface we have
\begin{equation} 
\int_V g_\alpha^{f,{\rm ext}}dV=-\int_S \sigma_{\alpha\beta}^0 dA_\beta
\end{equation}
Therefore, for any surface $S$ enclosing the particle in the fluid,
\begin{equation}
f_\alpha^{\rm tot}=\int_{V_p}(g_\alpha^{p,{\rm ext}}-g_\alpha^{f,{\rm ext}})dV+\int_S(\sigma_{\alpha\beta}
-\sigma^0_{\alpha\beta})dA_\beta=0
\end{equation}
 We thus find that
 the relative external body force
 \begin{equation}
 f_\alpha^p=\int_{V_p}(g_\alpha^{p,{\rm ext}}-g_\alpha^{f,{\rm ext}})dV
 \end{equation}
 which includes an Archimedian correction to the net body force, 
 is balanced by the stresses
 exerted by the perturbation of the flow field on the particle
 \begin{equation}
 f_\alpha^p=-\int_{S}(\sigma_{\alpha\beta}
-\sigma^0_{\alpha\beta})dA_\beta \label{fext}
 \end{equation}
 (for the choice $S=S_p$). 
 
 Eq. (\ref{fext}) is valid for arbitrary $S$ enclosing the particle.
  Therefore, for colloidal particles far from the boundaries of the fluid, the total momentum 
 flux through any closed surface is constant and equal to $f_\alpha^p$. 
 This implies that the stress field perturbation $\sigma_{\alpha\beta}-\sigma^0_{\alpha\beta}\sim 1/r^2$ 
 for large r, where $r$ is the distance from the particle center. 
 Since $\sigma_{\alpha\beta}-\sigma_{\alpha\beta}^0\sim (\partial_\alpha v_\beta+\partial_\beta v_\alpha)$, 
 the barycentric velocity perturbation scales as $\vert {\bf v}\vert \sim 1/r$ which implies 
 the existence of a Stokeslet. 
If the relative external body force $f_\alpha^p$ 
vanishes, but a relative external 
torque acts on the particle, a similar argument using angular moment conservation implies 
that in this case the stress scales as $ \sim 1/r^3$   and the velocity field like $\sim 1/r^2$. 

If no external  field acts on either 
the particle or the fluid, there is neither external body force no torque. 
The far field of the hydrodynamic flow 
is dominated by a force dipole or Stokes doublet with stress $\sim 1/r^3$ and velocity $\sim 1/r^2$. 
If the force dipole vanishes, %
the stress scales as $\sim 1/r^4$ and  the velocity as $\sim 1/r^3$.
As we have illustrated in this work, the absence of external forces and torques does not mean 
that there is no motion between particle and fluid.

\end{document}